\documentclass[aps,prl,nopacs,twocolumn,preprintnumbers,notitlepage,amsmath,amssymb,superscriptaddress]{revtex4-2}

\usepackage{graphicx}% Include figure files
\usepackage{dcolumn}% Align table columns on decimal point
\usepackage{bm}% bold math
\newcommand{\kcat}[0]{k_{\mathrm{cat}}}
\newcommand{\kspo}[0]{k_{\mathrm{spo}}}
\newcommand{\rcat}[0]{r_{\mathrm{cat}}}
\newcommand{\rspo}[0]{r_{\mathrm{spo}}}

\newcommand{\Rs}[0]{R_\mathrm{s}}
\newcommand{\Rp}[0]{R_\mathrm{p}}

\newcommand{\phie}[0]{\phi_{\mathrm{e}}}
\newcommand{\phis}[0]{\phi_{\mathrm{s}}}
\newcommand{\phip}[0]{\phi_{\mathrm{p}}}
\newcommand{\phisp}[0]{\phi_{\mathrm{s+p}}}
\newcommand{\phiw}[0]{\phi_{\mathrm{w}}}

\newcommand{\mue}[0]{\mu_{\mathrm{e}}}
\newcommand{\muf}[0]{\mu_{\mathrm{f}}}
\newcommand{\mus}[0]{\mu_{\mathrm{s}}}
\newcommand{\mup}[0]{\mu_{\mathrm{p}}}
\newcommand{\muw}[0]{\mu_{\mathrm{w}}}

\newcommand{\Des}[0]{D_{\mathrm{es}}}
\newcommand{\Dse}[0]{D_{\mathrm{se}}}
\newcommand{\Dep}[0]{D_{\mathrm{ep}}}
\newcommand{\Dpe}[0]{D_{\mathrm{pe}}}

\newcommand{\Dps}[0]{D_{\mathrm{ps}}}
\newcommand{\Dew}[0]{D_{\mathrm{ew}}}
\newcommand{\Dsw}[0]{D_{\mathrm{sw}}}
\newcommand{\Dpw}[0]{D_{\mathrm{pw}}}

\newcommand{\Mee}[0]{M_{\mathrm{ee}}}
\newcommand{\Mes}[0]{M_{\mathrm{es}}}
\newcommand{\Mep}[0]{M_{\mathrm{ep}}}
\newcommand{\Mse}[0]{M_{\mathrm{se}}}
\newcommand{\Mss}[0]{M_{\mathrm{ss}}}
\newcommand{\Msp}[0]{M_{\mathrm{sp}}}
\newcommand{\Mpe}[0]{M_{\mathrm{pe}}}
\newcommand{\Mps}[0]{M_{\mathrm{ps}}}
\newcommand{\Mpp}[0]{M_{\mathrm{pp}}}
\newcommand{\Mwp}[0]{M_{\mathrm{wp}}}
\newcommand{\Mws}[0]{M_{\mathrm{ws}}}
\newcommand{\Mwe}[0]{M_{\mathrm{we}}}

\newcommand{\vp}[0]{v_{\mathrm{p}}}
\newcommand{\vs}[0]{v_{\mathrm{s}}}
\newcommand{\ve}[0]{v_{\mathrm{e}}}
\newcommand{\vw}[0]{v_{\mathrm{w}}}

\newcommand{\Ree}[0]{R_{\mathrm{e}}}

\usepackage{amssymb}
\usepackage{amsmath}
\usepackage{color}
\usepackage[caption=false]{subfig}
\usepackage{mathtools,bm}
\usepackage{makecell}
\usepackage{enumerate}
\usepackage{mathtools}
\usepackage{stmaryrd}
\usepackage{enumitem}
\usepackage{textcomp}
\usepackage{gensymb}
\usepackage{notes2bib}

\begin{document}

\title{Catalysis-Induced Phase Separation and Autoregulation of Enzymatic Activity}

\author{Matthew W. Cotton}
\affiliation{Mathematical Institute, University of Oxford, Woodstock Road, Oxford, OX2 6GG, United Kingdom}
\affiliation{Department of Living Matter Physics, Max Planck Institute for Dynamics and Self-Organization, D-37077 G\"ottingen, Germany}

\author{Ramin Golestanian}\email{ramin.golestanian@ds.mpg.de}
\affiliation{Department of Living Matter Physics, Max Planck Institute for Dynamics and Self-Organization, D-37077 G\"ottingen, Germany}
\affiliation{Rudolf Peierls Centre for Theoretical Physics, University of Oxford, Oxford OX1 3PU, United Kingdom}

\author{Jaime Agudo-Canalejo}\email{jaime.agudo@ds.mpg.de}
\affiliation{Department of Living Matter Physics, Max Planck Institute for Dynamics and Self-Organization, D-37077 G\"ottingen, Germany}

\date{\today}

\begin{abstract}
We present a thermodynamically consistent model describing the dynamics of a multi-component mixture where one enzyme component catalyzes a reaction between other components. We find that the catalytic activity alone can induce phase separation for sufficiently active systems and large enzymes, without any equilibrium interactions between components. In the limit of fast reaction rates, binodal lines can be calculated using a mapping to an effective free energy. We also explain how this catalysis-induced phase separation (CIPS) can act to autoregulate the enzymatic activity, which points at the biological relevance of this phenomenon.
\end{abstract}
%<600 characters

\maketitle

\textit{Introduction.}---Liquid-liquid phase separation has emerged in recent years as a key principle governing intracellular organization \cite{banani2017biomolecular,shinLiquidPhaseCondensation2017}. It is generally believed that the main drivers of phase separation in such systems are the attractive equilibrium interactions between the different soluble components, which are needed to overcome the entropic costs associated with phase separation \cite{brangwynnePolymerPhysicsIntracellular2015,berryPhysicalPrinciplesIntracellular2018}. The emergence of condensates that are enriched or depleted in specific molecules can be designed by tuning these interactions \cite{mao2019phase,mao2020designing,jacobsSelfAssemblyBiomolecularCondensates2021,shrinivas2021phase}. On the other hand, it is clear that intracellular environments are far from being at thermodynamic equilibrium, and that the possible effects of non-equilibrium activity on phase separation need to be taken into consideration \cite{weber2019physics,kirschbaum2021controlling,liNonequilibriumPhaseSeparation2020,wittkowski2014scalar,tjhung2018cluster,saha2020scalar,you2020nonreciprocity}. In all of these studies, however, equilibrium interactions have remained the driving force for phase separation; non-equilibrium effects have entered only as additional chemical reactions that convert the phase-separating components into each other \cite{weber2019physics,kirschbaum2021controlling,liNonequilibriumPhaseSeparation2020} or, in a coarse-grained description, as gradient \cite{liNonequilibriumPhaseSeparation2020,wittkowski2014scalar,tjhung2018cluster} or non-reciprocal \cite{saha2020scalar,you2020nonreciprocity} terms that do not derive from a free energy. These non-equilibrium effects, albeit not driving the phase separation process, may for example affect the size distribution and coarsening dynamics of the resulting condensates, or lead to the formation of static and moving micropatterns.

Biomolecular condensates are often rich in enzymes that catalyze chemical reactions, in which case they are known as metabolons \cite{sweetloveRoleDynamicEnzyme2018}. Such enzyme-rich condensates can also be assembled \emph{in vitro} \cite{testa2021sustained}. The packing of enzymes in close proximity to each other can cause changes in metabolic and enzymatic rates when compared to a homogeneous system, for example by substrate channelling, where an intermediate product in a cascade reaction is passed on between enzymes \cite{poshyvailoDoesMetaboliteChanneling2017}, or by mechanical effects that alter the catalytic rate \cite{PhysRevLett.127.208103}.
Moreover, it has been suggested that biological systems can self-organize the cell cycle dynamics to lower the overall rate of metabolic activity and the ensuing free energy dissipation \cite{Niebel2019}. While the mechanisms underlying both the formation of enzyme-rich condensates and metabolic auto-regulation are currently not well understood \cite{sweetloveRoleDynamicEnzyme2018,Niebel2019}, it would be interesting to investigate whether such behaviours can generically emerge from spatial organization that arises from catalysis-induced non-equilibrium activity.

\begin{figure}[b]
	\begin{center}
		\includegraphics[width=1\linewidth]{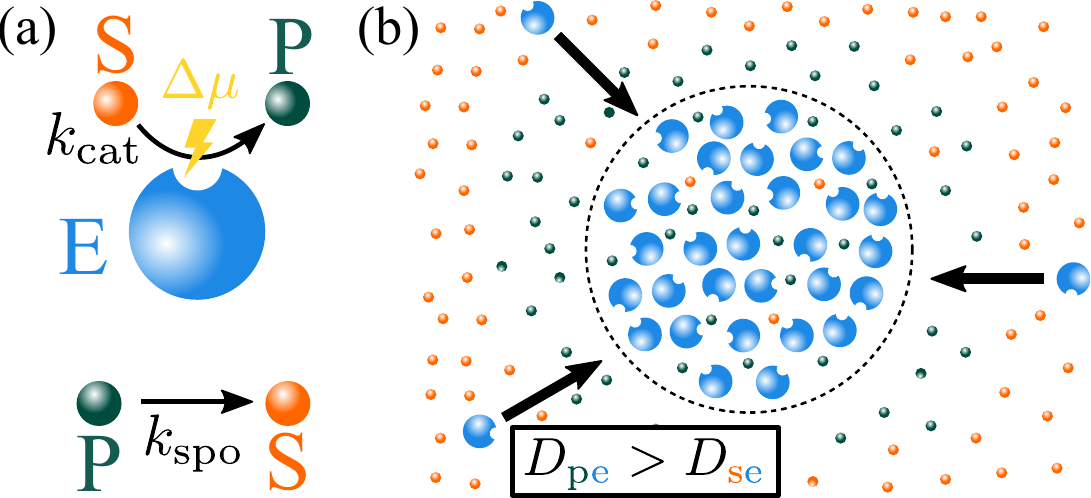}
		\caption{Processes leading to catalysis-induced phase separation (CIPS). 
		(a) Enzymes convert substrate into product by a fuelled catalytic reaction, while product turns into substrate spontaneously. (b) The catalyzed reaction creates gradients of substrate and product around enzyme-rich regions, which attract more enzymes when the off-diagonal transport coefficients coupling enzyme fluxes to product and substrate thermodynamic forces satisfy $\Dpe>\Dse$. \label{fig:cartoon-schematic}}
	\end{center}
\end{figure}

\begin{figure}%[t]
	\begin{center}
		\includegraphics[width=1\linewidth]{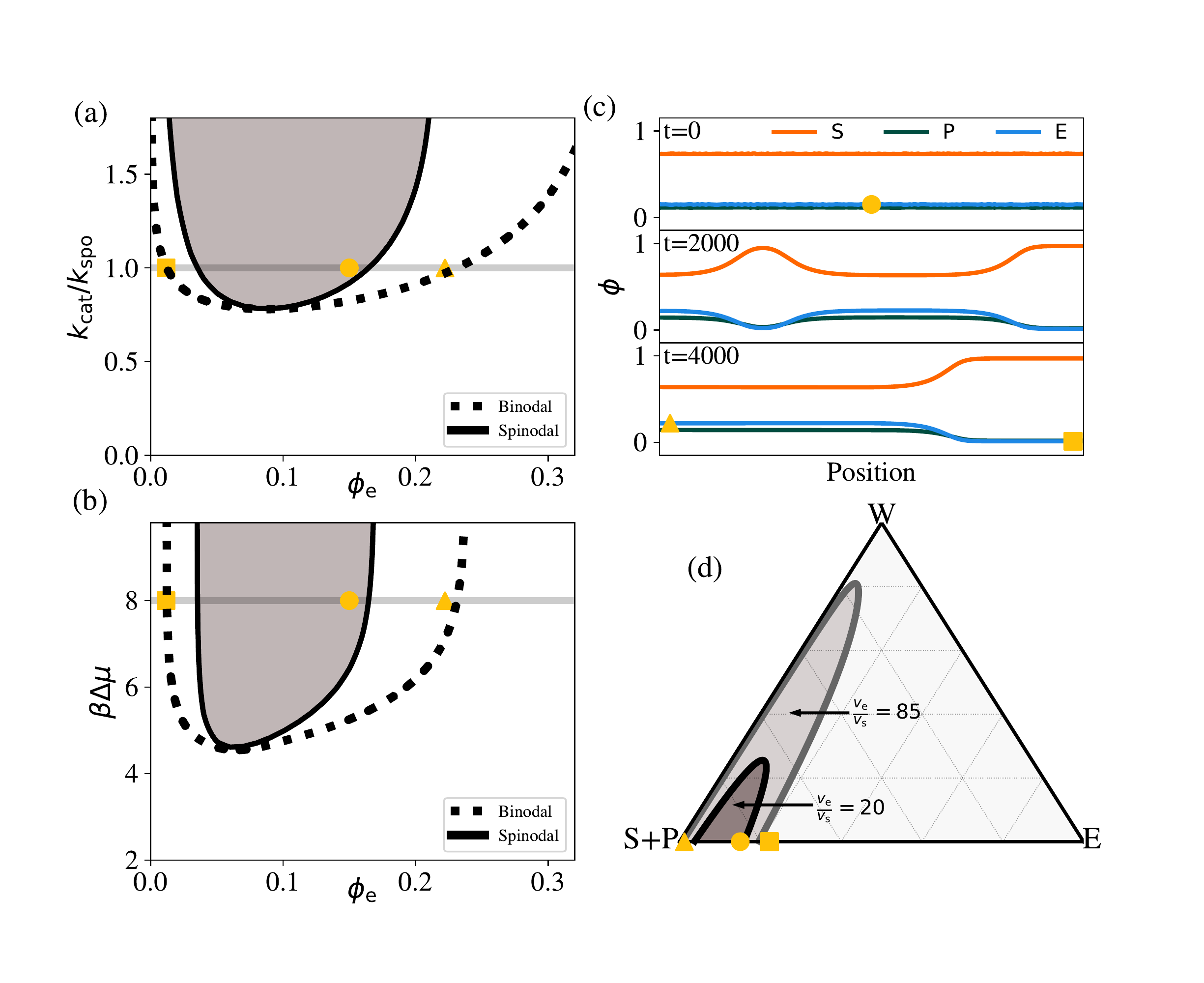}
		\caption{Phase behaviour and onset of CIPS. (a,b) Spinodal lines [from (\ref{eq:bi_stab})] and binodal lines (from the common tangent construction of $f_\mathrm{eff}$) for (a) varying $\kcat$ with $\Delta\mu=8 k_{\rm B} T$ and (b) varying $\Delta \mu$ with $\kcat/\kspo=1$. (c) Numerical simulations showing the evolution of a uniform steady state with $\phie=0.15$ into two phase separated regions. The circle, triangle and square identify the homogeneous steady state and the dense and dilute enzyme phases, respectively, and are plotted in all other panels for comparison. (d) Stability diagram of a mixture including a water component, for $\Delta\mu=8 k_{\rm B} T$ and $\kcat/\kspo=1$. The darker and lighter shaded regions mark the spinodal regions for $\ve/\vs=20$ [also used in (a--c)] and $\ve/\vs=85$, respectively. Additional system parameters in (a--d) are $\Delta \varepsilon=-5 k_{\rm B} T$, $\Dpe=4 \Dse$, and $\Dps=10 \Dse$; in (d) $\Dew=\Dsw=\Dpw=10 \Dse$ and $\vw=\vs$. \label{fig:bino_spino}}
	\end{center}
\end{figure}

Here, we propose a fundamentally new mechanism for the formation of enzyme-rich condensates, which does not rely on equilibrium attractive interactions between enzymes, but rather on effective interactions that arise purely as a consequence of their non-equilibrium catalytic activity (see Fig. \ref{fig:cartoon-schematic} for a schematic of the phenomenon and Fig. \ref{fig:bino_spino} for the corresponding phase diagrams). While effective interactions mediated by self-generated chemical gradients have been previously described in the context of phoretic active colloids or chemotactic microorganisms \cite{saha2014clusters, golestanian2019phoretic, agudo-canalejoActivePhaseSeparation2019, nasouri2020exact, keller1970initiation}, these were based on a microscopic and hydrodynamic description of individual colloid-colloid interactions. The theoretical framework presented here takes a complementary approach based on non-equilibrium thermodynamics and Flory-Huggins theory of suspensions, to manifestly connect the phenomenology to the existing studies on intracellular phase separation \cite{brangwynnePolymerPhysicsIntracellular2015, berryPhysicalPrinciplesIntracellular2018, mao2019phase,mao2020designing, jacobsSelfAssemblyBiomolecularCondensates2021, shrinivas2021phase, weber2019physics, kirschbaum2021controlling}.
We find that this catalysis-induced phase separation (CIPS) can be described by a mapping to an effective free energy, and thus shows equilibrium features such as the existence of binodal and spinodal lines which meet at a critical point. Moreover, we show that phase separation in this model, which is itself induced by catalysis, generically leads to a decrease in the overall catalytic activity of the system, thus providing a mechanism for the autoregulation of catalytic activity.

\textit{Model.}---We consider an incompressible fluid with $N$ components described by the volume fractions $\phi_i(\bm{r},t)$, each corresponding to individual molecules of volume $v_i$ on the microscopic scale. The Flory-Huggins theory of suspensions gives the free energy of the system as $F = \int \mathrm{d}\bm{r} f_\mathrm{FH}$, with the free energy density
\begin{math}%{equation}
    f_\mathrm{FH}(\{\phi_i\}) = \sum_{i=1}^{N} \frac{1}{v_i} \big[\varepsilon_i\phi_i + k_\mathrm{B}T \phi_i \big(\log\phi_i-1\big) \big], 
    \label{eq:fh_gen}
\end{math} %{equation}
where $\varepsilon_i$ is the enthalpy of component $i$. Importantly, we do not include any interaction terms in the free energy, in particular $f_\mathrm{FH}$ does not contain terms of the usual form $\chi_{ij}\phi_i \phi_j$. This implies that phase separation in this system would be impossible at equilibrium. We denote $\beta \equiv (k_\mathrm{B}T)^{-1}$. Each $\phi_i$ is governed by conserved dynamics $\dot{\phi}_i +\bm{\nabla}\cdot\bm{J}_i=0$ driven by thermodynamic fluxes $\bm{J}_i = - \sum_{j=1}^N M_{ij}\bm{\nabla}\mu_j$, where $\mu_j=v_j\frac{\delta F}{\delta \phi_j}=\varepsilon_j+k_\mathrm{B}T \log\phi_j$ is the chemical potential of component $j$, and $M_{ij}$ is a mobility matrix \cite{grootNonequilibriumThermodynamics1984}. Incompressibility of the suspension requires $\sum_{i=1}^N\phi_i=1$, which implies (via the dynamical equations)
that the mobilities must satisfy $\sum_i M_{ij} = 0$ \cite{kehr1989mobility}. The Onsager reciprocal relations further constrain the form of the mobilities, namely $v_j M_{ij} = v_i M_{ji}$ \cite{grootNonequilibriumThermodynamics1984}. These constraints mean that a system of $N$ components has $N(N-1)/2$ free mobilities. In the following, we assume the common form of $M_{ij} = -\beta D_{ij}\phi_i\phi_j$ for $i \neq j$, where the constraints just described imply $M_{jj}= - \sum_{i\neq j} M_{ij}$ and $v_j D_{ij} = v_i D_{ji}$ \cite{KRAMER1984473,kehr1989mobility,mao2020designing,bo2021}. We note that the transport coefficients $D_{ij}$ determine the rate at which the components respond to local effective concentration gradients and exchange positions, and as such are inherently related to the phenomena of diffusiophoresis, cross-diffusion and Maxwell-Stefan diffusion \cite{suppmat}.

We make the model active by allowing non-equilibrium (fuelled) conversion between two components, substrate (S) and product (P), catalyzed by an enzyme (E). This can be described by the reaction E+S+F $\rightleftharpoons$ E+P+W, where F and W represent fuel and waste molecules, respectively. We do not model the dynamics of the fuel and waste here, but assume that the system is in contact with a reservoir that maintains constant chemical potentials, $\muf$ and $\muw$, and define $\Delta\mu \equiv \muf-\muw$. Alternatively, $\Delta \mu$ could represent the energy transferred by a photon in a light-activated catalytic reaction. We further allow for spontaneous conversion between S and P, corresponding to the reaction S $\rightleftharpoons$ P. Note that incompressibility implies $\vp = \vs$. Using the definition $\Delta \varepsilon \equiv \varepsilon_\mathrm{s}-\varepsilon_\mathrm{p}$, we can write the net rate of the spontaneous reaction as
\begin{equation}
    \rspo = \rspo^{\mathrm{S}\to\mathrm{P}} - \rspo^{\mathrm{P}\to\mathrm{S}} = \kspo[e^{\beta \Delta \varepsilon}\phis - \phip],
\end{equation}
which entails detailed balance with $\rspo^{\mathrm{S}\to\mathrm{P}}/\rspo^{\mathrm{P}\to\mathrm{S}} =\mathrm{exp}[\beta (\mus-\mup)]$. The catalyzed reaction rate will have a similar functional form with an additional dependence on $\phie$, namely 
\begin{equation}
    \rcat = \rcat^{\mathrm{S}\to\mathrm{P}} - \rcat^{\mathrm{P}\to\mathrm{S}} = \kcat\phie[\phis-\phip e^{-\beta(\Delta \varepsilon+ \Delta\mu)}],
    \label{r_cat}
\end{equation}
which also entails detailed balance with $\rcat^{\mathrm{S}\to\mathrm{P}}/\rcat^{\mathrm{P}\to\mathrm{S}}=\mathrm{exp}[\beta (\mus-\mup+\Delta\mu)]$. We will typically take $\Delta \varepsilon<0$ and $\Delta \varepsilon + \Delta \mu>0$, so that the spontaneous and catalyzed reactions run preferentially in the P$\to$S and S$\to$P directions, respectively; see Fig.~\ref{fig:cartoon-schematic}. Combining the conserved dynamics with the reaction terms and defining $R\equiv\rspo+\rcat$ results in the evolution equations for the three-component system
\begin{align}
    \dot{\phie} &= \bm{\nabla} \cdot \big(\Mee\bm{\nabla}\mue + \Mes\bm{\nabla}\mus + \Mep\bm{\nabla}\mup\big), \label{evoE}\\
    \dot{\phis} &= \bm{\nabla} \cdot \big(\Mse\bm{\nabla}\mue + \Mss\bm{\nabla}\mus + \Msp\bm{\nabla}\mup\big) - R, \label{evoS}\\
    \dot{\phip} &= \bm{\nabla} \cdot \big(\Mpe\bm{\nabla}\mue + \Mps\bm{\nabla}\mus + \Mpp\bm{\nabla}\mup\big) + R. \label{evoP}
\end{align}

\textit{Steady-state and stability.}---The minimal model [Eqs. (\ref{evoE})--(\ref{evoP})] has a homogeneous steady-state solution when $R=0$, which is given by any $\phie^*$ as well as
\begin{equation}
    \phis^* = \phisp^*\frac{\kspo + \kcat\phie^*e^{-\beta(\Delta \varepsilon + \Delta\mu)}}{\kspo + \kcat\phie^*e^{-\beta (\Delta \varepsilon + \Delta\mu)}+\kspo e^{\beta \Delta \varepsilon} + \kcat\phie^*}
    \label{sstar}
\end{equation}
with $\phisp^*=1-\phie^*$ and $\phip^*=\phisp^*-\phis^*$.

    We can study the linear stability of this homogeneous steady-state by considering a small perturbation $\phi_i(\bm{r},t) = \phi_i^* + \delta\phi_i(\bm{r}, t)$. We find that the steady-state undergoes an instability at the longest wavelengths 
provided the following condition holds
\begin{equation}
    \frac{1}{\ve \phie^*}+\frac{1}{\vs(1-\phie^*)} < \frac{\kspo \kcat   (1-e^{-\beta \Delta \mu})(\Dpe-\Dse)}{\vs( \Rs + \Rp )(\Dpe \Rs+\Dse \Rp)},
    \label{eq:bi_stab}
\end{equation}
where we have defined $\Rs \equiv \kspo e^{\beta \Delta \varepsilon} + \kcat\phie^*$ and $\Rp \equiv \kspo + \kcat\phie^*e^{-\beta (\Delta \varepsilon+\Delta \mu)}$ \cite{suppmat}. Since the left hand side of (\ref{eq:bi_stab}) is always positive, an instability is possible only if the right hand side is positive as well. The sign of the right hand side is controlled by that of $(1-e^{-\beta\Delta \mu})(\Dpe-\Dse)$, which has several implications. First, an equilibrium system with $\Delta\mu=0$ is always stable. Second, for a catalytic reaction favouring product formation with $\Delta \mu>0$, an instability is possible only if $\Dpe>\Dse$. Third, if $\Dpe=\Dse$ the system is always stable. Intuitively, the instability arises from opposing gradients of substrate and product generated around an enzyme-rich region when $\Delta \mu > 0$, coupled to an unequal response of the enzyme to gradients of substrate and product when $\Dpe>\Dse$, resulting in effective enzyme-enzyme attractive interactions and further aggregation. Interestingly, the instability is favoured when $\ve \gg \vs$, which happens to correspond to the typical relative sizes of enzymes and substrates in biological systems. In Fig.~\ref{fig:bino_spino}(a,b), the unstable region delimited by Eq. (\ref{eq:bi_stab}) is shown as a function of the catalytic rate $\kcat$ and the non-equilibrium drive $\Delta \mu$.

Numerical solution of the evolution equations (\ref{evoE})--(\ref{evoP}) confirms the existence of this instability. We initialize a 1D system
with small number-conserving random variations around $\phi_i^*$. When the system is unstable, regions of high and low enzyme concentrations develop spontaneously [see Fig.~\ref{fig:bino_spino}(c)] and coarsen over time, ultimately resulting in two distinct phase-separated domains. Moreover, varying the amount of enzyme in the system only changes the relative size of the high and low concentration domains, without affecting the concentration values in the two domains, which suggests the existence of a binodal line, as in equilibrium phase separation. We observed this behavior for all parameters which we simulated ($\kcat/\kspo\approx 1 \text{--} 100$, $\ve/\vs\approx 2 \text{--} 100$, $-\Delta \varepsilon\approx 5 \text{--} 30 k_{\rm B} T$, $\Delta \mu\approx5 \text{--} 50k_{\rm B} T$,  $\Dpe/\Dse\approx1 \text{--} 100$). The observation of macroscopic phase separation, rather than pattern formation or microphase separation, is further supported by the linear stability analysis showing an instability at the largest wavelengths ($q^2 \to 0$), rather than at finite wavelengths. We also note that studies of two-component mass-conserving reaction-diffusion systems, which have significant parallels to the model studied here \cite{suppmat}, have shown that these systems exhibit uninterrupted coarsening leading to macrophase separation at long times \cite{brauns2020phase,brauns2021wavelength}.

\textit{Effective free energy and binodal.}---In the macroscopic limit, we expect the substrate-product equilibrium in the bulk of each phase to be governed by the reaction terms that act locally, rather than by spatial diffusion. This implies that the substrate and product concentrations are enslaved to the enzyme concentration by $\phis \approx \phis^*(\phie)$ and $\phip \approx \phip^*(\phie)$, with the functions defined in (\ref{sstar}). Substituting these expressions into (\ref{evoE}), we can recast the dynamics of the enzyme as $\dot{\phie}\approx \bm{\nabla} \cdot (\Mee\bm{\nabla}\mu_\mathrm{eff})$ with an effective chemical potential for the enzyme
\begin{equation}
    \frac{\mu_\mathrm{eff}(\phie)}{k_{\rm B} T} =  \log \phie - \frac{\ve}{\vs} \log[\Dse \phis^* (\phie) + \Dpe \phip^*(\phie)].
\end{equation}
We can also identify an effective free energy density $f_\mathrm{eff}(\phie)$, such that $\mu_\mathrm{eff}=\ve\frac{\mathrm{d} f_\mathrm{eff}}{\mathrm{d} \phie}$, which can be explicitly calculated by direct integration \cite{suppmat}. 
By employing the common-tangent construction in unstable cases, we can identify two coexisting phases and define the binodal lines, which show good agreement with our numerical results and meet the spinodal line at a critical point; see Fig.~\ref{fig:bino_spino}(a,b).

\textit{The role of solvent.}---While we have so far considered an enzyme-substrate-product system for simplicity, we observe that an instability can also occur in the presence of an additional solvent, typically water, in which these components will be dissolved. We can add a fourth component of volume fraction $\phiw$ to the dynamics and study the stability of the homogeneous steady-state \cite{suppmat}. 
We find that the uniform steady-state can be unstable even when all the solute components (enzyme, substrate, and product) are in dilute conditions, as shown in Fig.~\ref{fig:bino_spino}(d). This demonstrates the wide reach of this work and its potential application to realistic systems. For the remainder of this Letter, however, we focus on the simpler $\phiw = 0$ case.

\begin{figure}[t]
	\begin{center}
		\includegraphics[width=1\linewidth]{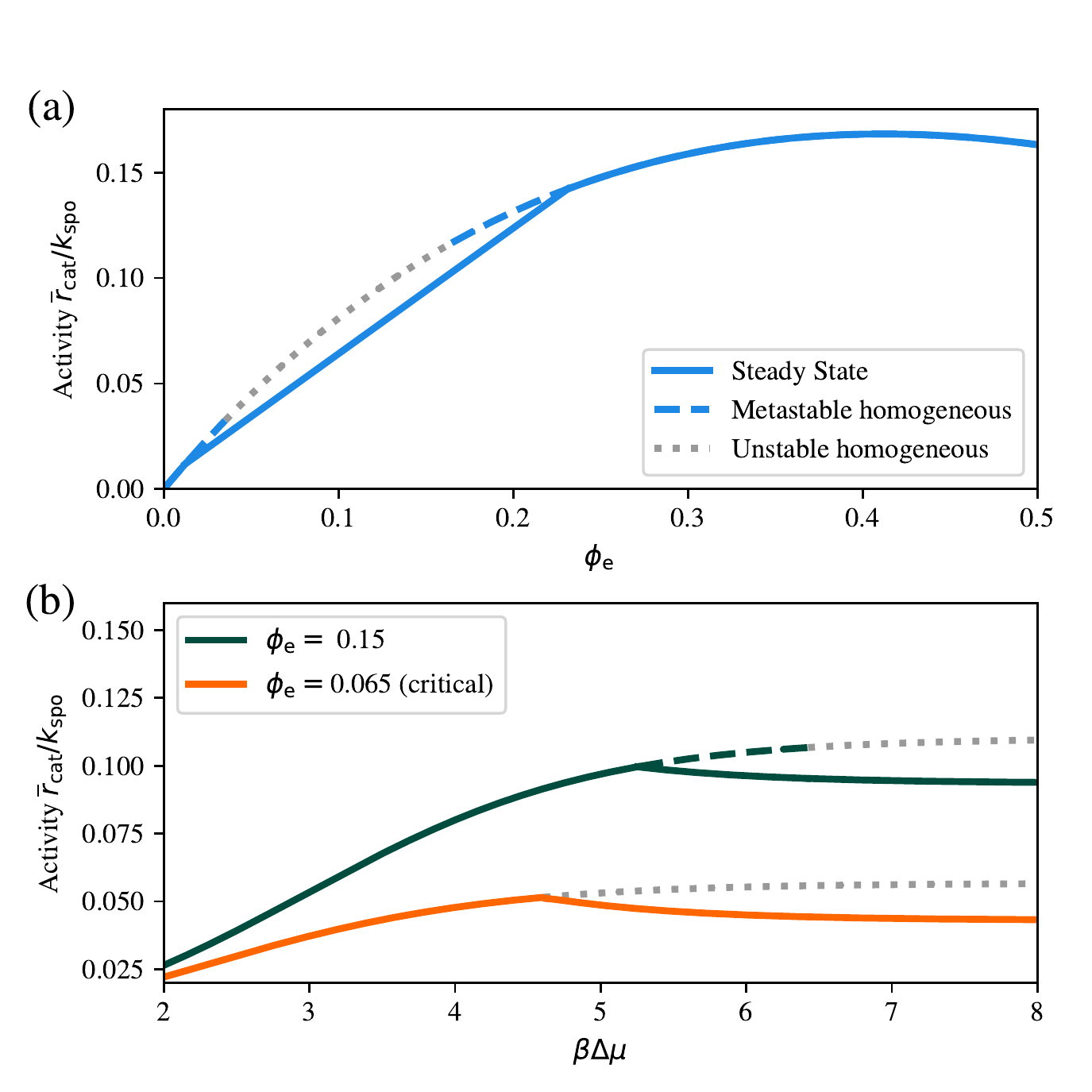}
		\caption{Effect of CIPS on catalytic activity. (a) Activity as a function of the initial $\phie$. In the phase separated state, the activity is a linear combination of the activity of the homogeneous states on either side of the binodal, which due to convexity is always smaller than that of the homogeneous state. (b) Evolution of activity with increasing $\Delta \mu$. As the system phase separates, the overall catalytic rate is reduced. When $\phie=0.065$, the system passes through the critical point and there is no metastable homogeneous branch. System parameters are as in Fig.~\ref{fig:bino_spino}. \label{fig:activity}}
	\end{center}
\end{figure}

\textit{Enzymatic autoregulation.}---A biologically pertinent question is what happens to the enzymatic activity when the system phase separates. The average rate of catalysis in a region of size $L$ is given by $\bar{r}_\mathrm{cat} = \frac{1}{L}\int_0^L r_\mathrm{cat} \mathrm{d}x$ in a simple 1D case. In a homogeneous state, $r_\mathrm{cat}$ will be constant throughout the system and, using Eq. (\ref{r_cat}), will go as $\bar{r}_\mathrm{cat}^{\mathrm(h)}\sim\phie(1-\phie)$ which is a concave function of $\phie$. In a phase separated state, $\bar{r}_\mathrm{cat}$ is a weighted average of the catalytic rates in each phase, with the weights determined by the lever rule. Due to the concavity of $\bar{r}_\mathrm{cat}^{\mathrm(h)}$, we find that the catalytic rate in the phase separated state is always smaller than in the homogeneous state; see Fig.~\ref{fig:activity}(a). We observe a similar behaviour when we vary a control parameter such as $\Delta\mu$, which is controlled by the concentration of the fuel molecules in an experiment; see Fig.~\ref{fig:activity}(b). In the homogeneous phase, $\bar{r}_\mathrm{cat}$ initially rises and then saturates with increasing $\Delta\mu$. The phase separation reduces $\bar{r}_\mathrm{cat}$ in the whole system and leads to saturation at a lower activity. Through this mechanism, CIPS can act to autoregulate the enzymatic activity of the mixture: once the activity reaches a threshold, the system phase separates and gives rise to a reduced overall catalytic rate. A similar saturation effect is seen when other system parameters, such as $\kcat$, are varied causing the system to phase separate. 

\textit{Discussion.}---Using a thermodynamically-consistent description of a multicomponent fluid based on linear response theory constructed from a Flory-Huggins free energy, we have identified a new, purely non-equilibrium mechanism for phase separation as a consequence of the catalytic, fuelled conversion between two components (substrate and product) by a third component (enzyme). Besides the catalytic activity, a necessary ingredient for catalysis-induced phase separation (CIPS) is an asymmetry in the off-diagonal response coefficients (mobilities) that couple enzyme-substrate and enzyme-product thermodynamic forces and fluxes in the non-equilibrium conserved dynamics. Using a mapping of the three-component system to a single-component system with an effective free energy, equilibrium-like features of CIPS such as binodal lines were obtained.

We argue that the substrate-vs-product mobility asymmetry required for CIPS to operate can plausibly exist in realistic systems. For a typical biological catalytic process, we expect both the spontaneous and catalyzed reactions to be strongly driven, and the enzyme protein to be much larger than the small molecular substrate and product. In this biologically realistic limit, we find \cite{suppmat} that CIPS occurs at low enzyme concentrations whenever ${(\Dpe-\Dse)}/{\Dse} > {\kspo}/{\kcat}$. Given that the kinetics of catalyzed reactions are generally much faster than those of spontaneous ones (reduced energy barrier, with $\kcat\gg\kspo$), this implies that the threshold mobility asymmetry required for CIPS can become vanishingly small. While measurements of the off-diagonal Onsager mobilities for biologically relevant enzyme-substrate-product systems do not exist at present to the best of our knowledge, measurements of the functionally equivalent (see \cite{suppmat}) Maxwell-Stefan diffusivities of various multicomponent mixtures suggest that even small changes in molecular structure (e.g.~shape, polarity, etc.) of the mixture components can result in substantial changes to the mobilities \cite{taylor1993multicomponent,guevara2016mutual,guevara2018interplay,ramm2021diffusiophoretic,vanag2009cross}.

The mechanism behind CIPS is reminiscent of mechanisms for chemotactic or phoretic aggregation previously described in the literature in the context of interacting microorganisms or catalytically-active colloids \cite{saha2014clusters,golestanian2019phoretic,agudo-canalejoActivePhaseSeparation2019,keller1970initiation}. However, these studies were based on microscopic descriptions of the chemotactic or phoretic response, typically valid only under dilute conditions. We expect that such microscopic descriptions and the thermodynamic-phenomenological description presented here are two sides of the same coin, the former being applicable arbitrarily far from equilibrium in dilute conditions, the latter near equilibrium at arbitrary densities.
Indeed, a connection can be formally established between the off-diagonal Onsager mobilities and phoretic mobilities \cite{golestanian2019phoretic,anderson1989colloid} or, equivalently, the Fickian cross-diffusion coefficients \cite{vanag2009cross} (see \cite{suppmat} for details). The existing experimental observations \cite{agudo2018enhanced,zhang2021chemically} of unequal response of enzymes to gradients of substrate and product thus further corroborate the assertion that an asymmetry may generically exist between the enzyme-substrate and enzyme-product Onsager mobilities.

When the enzymatic activity in the homogeneous system is increased beyond a critical threshold, for example via external factors such as the availability of fuel molecules, the system phase separates, causing the overall enzymatic activity of the system to suddenly decrease and then plateau.
In multi-step metabolic pathways, the production of intermediate metabolites is known to regulate other reactions in the networks and thus act as a feedback mechanism that inhibits overall metabolic activity \cite{o2012dynamic,alam2017self}. CIPS provides a novel mechanism for this complex control of metabolism which, somewhat uniquely, autoregulates a single-step catalytic reaction and provides a simpler mechanism, potentially more amenable to fine-tuned synthetic control.
It remains to be seen how CIPS affects catalytic activity in multi-step metabolic reactions involving several distinct enzymes. We speculate that, in a system with several enzyme components, CIPS may allow for colocalization of distinct enzymes within the same aggregate, allowing for substrate channelling as in cellular metabolons \cite{poshyvailoDoesMetaboliteChanneling2017,sweetloveRoleDynamicEnzyme2018}. Indeed, we previously showed that this behaviour is possible in mixtures of phoretic active colloids \cite{agudo-canalejoActivePhaseSeparation2019}.

Owing to its nonequilibrium nature, CIPS results in phase separated states with non-vanishing fluxes, and is distinct from equilibrium mechanisms for phase separation. The latter rely on the presence of interaction terms (e.g.~$\chi_{ij}\phi_i\phi_j$ and $\kappa_{ij}\bm{\nabla} \phi_i \cdot \bm{\nabla} \phi_j$) in the free energy density $f_\mathrm{FH}$, which may be of enthalpic (temperature-independent) or entropic (temperature-dependent) origin. In particular, despite also requiring a size difference between components, CIPS is distinct from the entropic phase separation induced by depletion effects that is observed in binary hard-core mixtures \cite{frenkel1992phase}, which results in equilibrium phase separated states with vanishing fluxes.
Future work may explore the competition or cooperation between equilibrium interactions and non-equilibrium catalytic effective interactions in phase separation.
In particular, we note that we have focused here on effective interactions that are attractive, i.e.~those with $(1-e^{-\beta \Delta \mu})(\Dpe-\Dse)>0$ so that the right hand side of (\ref{eq:bi_stab}) is positive. One may also consider repulsive effective interactions, with $(1-e^{-\beta \Delta \mu})(\Dpe-\Dse)<0$. In this case, we expect that an enzyme-rich condensate held together by equilibrium interactions may be {\it dissolved} by sufficiently strong non-equilibrium catalytic activity. This further highlights how the mechanism we have uncovered goes well beyond the prototypical example presented here, and may prove an important player in the description of phase separation in out-of-equilibrium systems.

This work has received support from the Max Planck School Matter to Life and the MaxSynBio Consortium, which are jointly funded by the Federal Ministry of Education and Research (BMBF) of Germany, and the Max Planck Society. M.C.~was supported by funding from the Biotechnology and Biological Sciences Research Council (UKRI-BBSRC) [grant number BB/T008784/1].

\bibliography{enz_agg.bib}

\newpage
\widetext
\begin{center}
	\textbf{\large Supplemental Material:\\Catalysis-Induced Phase Separation and Autoregulation of Enzymatic Activity}
\end{center}
%%%%%%%%%% Merge with supplemental materials %%%%%%%%%%
%%%%%%%%%% Prefix a "S" to all equations, figures, tables and reset the counter %%%%%%%%%%
\setcounter{equation}{0}
\setcounter{figure}{0}
\setcounter{table}{0}
\setcounter{page}{1}
\makeatletter
\renewcommand{\theequation}{S\arabic{equation}}
\renewcommand{\thefigure}{S\arabic{figure}}

\section{Stability of the three-component system}
\subsection{Condition of instability}
We consider the stability of the homogeneous steady state of a three component mixture described by volume fractions $\phie^*$, $\phis^*$ and $\phip^*$ which satisfy $R=0$. We consider a small perturbation from this uniform steady state $\phi_i(\bm{r},t) = \phi_i^* + \delta\phi_i(\bm{r},t)$ giving $\mu_i(\bm{r},t) = \mu_i^* + \delta\mu_i(\bm{r},t)$ and define mobilities at the homogeneous steady state $M^*_{ij} = M_{ij}(\phie^*, \phis^*, \phip^*)$. To first order in the perturbation $\delta\mu(\bm{r},t) = \frac{\delta\phi_i(\bm{r},t)}{\phi_i^*}$ and thus $\bm{\nabla}\mu_i(\bm{r},t) = \frac{1}{\phi_i^*}\bm{\nabla}\delta\phi_i(\bm{r},t)$, and moreover $R = \Rs \delta \phis - \Rp \delta \phip + \Ree \delta \phie$ where $\Rs = \kspo e^{\beta \Delta \varepsilon} + \kcat\phie^*$ and $\Rp = \kspo + \kcat\phie^*e^{-\beta (\Delta \varepsilon+\Delta \mu)}$ as defined in the main text, and
\begin{equation}
	\Ree \equiv k_\mathrm{cat} [ \phis^* - \phip^* e^{-\beta (\Delta \varepsilon + \Delta \mu)}] = \phi^*_\mathrm{s+p} \frac{k_\mathrm{cat} k_\mathrm{spo} (1-e^{-\beta \Delta \mu})}{k_\mathrm{spo} + k_\mathrm{cat} \phie^* e^{-\beta (\Delta \varepsilon + \Delta \mu)} + k_\mathrm{spo} e^{\beta \Delta \varepsilon} + k_\mathrm{cat} \phie^* }.
\end{equation}
The governing equations for the perturbations $\delta\phi_i (\bm{q},t) $ in Fourier space are found as follows\\
\noindent
\makebox[\textwidth]{\parbox{1.3\textwidth}{%
		\begin{equation}
			\begin{pmatrix}
				\dot{\delta\phie}\\
				\dot{\delta\phis}\\
				\dot{\delta\phip}\\
			\end{pmatrix} =
			\underbrace{
				\begin{pmatrix}
					(\Mse^*+\Mpe^*)\frac{\bm{q}^2}{\phie^*} & -\Mse^*\frac{\ve}{\vs}\frac{\bm{q}^2}{\phis^*} & -\Mpe^*\frac{\ve}{\vs}\frac{\bm{q}^2}{\phip^*}\\
					-\Mse^*\frac{\bm{q}^2}{\phie^*} - \Ree  & \big(\Msp^* + \Mse^*\frac{\ve}{\vs}\big)\frac{\bm{q}^2}{\phis^*} - \Rs  & -\Msp^*\frac{\bm{q}^2}{\phip^*} + \Rp \\
					-\Mpe^*\frac{\bm{q}^2}{\phie^*}  + \Ree  & -\Msp^*\frac{\bm{q}^2}{\phis^*} + \Rs  & \big(\Mpe^*\frac{\ve}{\vs}+\Msp^*\big)\frac{\bm{q}^2}{\phip^*} - \Rp \\
				\end{pmatrix}
			}_{\cal C}
			\begin{pmatrix}
				\delta\phie\\
				\delta\phis\\
				\delta\phip\\
			\end{pmatrix}.
\end{equation}}}
In this formulation, we can explicitly see the incompressibility being imposed as the columns of $\cal C$ sum to zero.
We now eliminate one of the components, e.g. $\delta\phip = -\delta\phie - \delta\phis$, and define a $2\times2$ matrix, $\cal K$, describing the evolution of the remaining components. This is obtained from $\cal C$ by subtracting the last column from the other two (i.e. ${\cal K}_{ij} = {\cal C}_{ij} - {\cal C}_{i3}$ for $i, j \in {1, 2}$), giving
\newline
\noindent
\makebox[\textwidth]{\parbox{1.3\textwidth}{%
		\begin{equation}
			\begin{pmatrix}
				\dot{\delta\phie}\\
				\dot{\delta\phis}\\
			\end{pmatrix} =
			\underbrace{
				\begin{pmatrix}
					(\Mse^*+\Mpe^*)\frac{\bm{q}^2}{\phie^*} +\Mpe^*\frac{\ve}{\vs}\frac{\bm{q}^2}{\phip^*}& -\Mse^*\frac{\ve}{\vs}\frac{\bm{q}^2}{\phis^*} + \Mpe^*\frac{\ve}{\vs}\frac{\bm{q}^2}{\phip^*}\\
					-\Mse^*\frac{\bm{q}^2}{\phie^*} - \Ree  + \Msp^*\frac{\bm{q}^2}{\phip^*} - \Rp  & \big(\Msp^* + \Mse^*\frac{\ve}{\vs}\big)\frac{\bm{q}^2}{\phis^*} - \Rs  + \Msp^*\frac{\bm{q}^2}{\phip^*} - \Rp \\
				\end{pmatrix}
			}_{\cal K}
			\begin{pmatrix}
				\delta\phie\\
				\delta\phis\\
			\end{pmatrix}.
\end{equation}}}
The eigenvalues of ${\cal K}$ determine the stability of the steady state. The system is stable if and only if all eigenvalues of ${\cal K}$ are negative. The Onsager relations and the non-negativity of $R_i$ mean ${\cal K}_{1,1}$ and ${\cal K}_{2,2}$ are always negative, so $\text{tr}({\cal K}) < 0$ and there is therefore one positive eigenvalue if $\det({\cal K})<0$. The instability condition then becomes
\begin{equation}
	\begin{split}
		\Bigg((\Mse^*+\Mpe^*)&\frac{\bm{q}^2}{\phie^*} + \Mpe^*\frac{\ve}{\vs}\frac{\bm{q}^2}{\phip^*}\Bigg)\Bigg(\big(\Msp^* + \Mse^*\frac{\ve}{\vs}\big)\frac{\bm{q}^2}{\phis^*} - \Rs  + \Msp^*\frac{\bm{q}^2}{\phip^*} - \Rp \Bigg) \\
		&< \Bigg(-\Mse^*\frac{\ve}{\vs}\frac{\bm{q}^2}{\phis^*} + \Mpe^*\frac{\ve}{\vs}\frac{\bm{q}^2}{\phip^*}\Bigg)\Bigg(-\Mse^*\frac{\bm{q}^2}{\phie^*} - \Ree  + \Msp^*\frac{\bm{q}^2}{\phip^*} - \Rp \Bigg).
	\end{split}
	\label{hi_ord_ins}
\end{equation}
When $\bm{q}$ is very large, we expect the system to be stable as in this regime the evolution is dominated by diffusive relaxation. For $\bm{q} = 0$, the system is at critical stability due to global conservation, as this corresponds to a uniform change in $\phi_i$. As such, since this condition only depends on $\bm{q}^2$ and $\bm{q}^4$, the instability is satisfied if and only if it is satisfied for very small $\bm{q}^2$. With this in mind we take Eq. (\ref{hi_ord_ins}) to order $\sim\bm{q}^2$ which gives
\begin{align}
	\frac{1}{\ve\phie^*}+\frac{1}{\vs(1-\phie^*)} < \frac{\Ree }{\vs(1-\phie^*)}\bigg(\frac{\gamma_\mathrm{p}}{\Rs }-\frac{\gamma_\mathrm{s}}{\Rp }\bigg)
	\label{eq_form_cond}
\end{align}
where we have defined
\begin{equation}
	\gamma_\mathrm{s} \equiv \frac{\Mse^*}{\Mse^*+\Mpe^*} \quad \text{   and    } \quad \gamma_\mathrm{p} \equiv \frac{\Mpe^*}{\Mse^*+\Mpe^*}.
\end{equation}
This condition can also be written to constrain the relative volumes in the system
\begin{equation}
	\frac{\vs}{\ve} < \frac{\phie^*}{1-\phie^*}\Bigg[\Ree \bigg(\frac{\gamma_\mathrm{p}}{\Rs }-\frac{\gamma_\mathrm{s}}{\Rp }\bigg) - 1\Bigg],
	\label{vol_ins}
\end{equation}
which shows that the instability is favoured by larger enzymes, as stated in the main text. Assuming the following form for the mobility $\Mse = -\beta \Dse\phie\phis$ and $\Mpe = -\beta \Dpe\phie\phip$, gives
\begin{equation}
	\frac{1}{\ve\phie^*}+\frac{1}{\vs(1-\phie^*)} < \frac{\Ree }{\vs(1-\phie^*)}\frac{\Dpe-\Dse}{\Dpe\Rs +\Dse\Rp }. \label{eq:inst}
\end{equation}
which corresponds to the instability condition given in the main text. In calculating this condition, we eliminated the product component from the system and determined the stability by considering the other two components. Eliminating either of the other components gives the same result.

\subsection{Fast dynamics}
In deriving the stability condition for the minimal active mixture, we note that the eigenvalues of a $2\times2$ matrix can be given in terms of its trace, $T$ and determinant, $D$, as $\lambda_{\pm} = \frac{1}{2}(T \pm \sqrt{T+4D})$. Alternatively, given the larger size of the enzymes, we could have made the assumption of fast substrate and product dynamics in comparison to the enzyme dynamics. Assuming that product and substrate equilibrate quickly, we can solve for $\dot{\delta\phis}=0$ to find a relationship between  $\delta\phis$ and $\delta\phie$. This gives 
\begin{equation}
	\dot{\delta\phis} = {\cal K}_{21}\delta\phie + {\cal K}_{22}\delta\phis = 0 \implies \delta\phis = -\frac{{\cal K}_{21}}{{\cal K}_{22}}\delta\phie \implies \dot{\delta\phie} = {\cal K}_{11}\delta\phie  - {\cal K}_{12}\frac{{\cal K}_{21}}{{\cal K}_{22}}\delta\phie
\end{equation}
and the instability condition then becomes
\begin{equation}
	{\cal K}_{11} - {\cal K}_{12}\frac{{\cal K}_{21}}{{\cal K}_{22}} > 0.
\end{equation}
Looking at the form of ${\cal K}_{22}$, in the small $\bm{q}$ limit, ${\cal K}_{22}=-(\Rs +\Rp ) < 0 $, which gives
\begin{equation}
	{\cal K}_{11}{\cal K}_{22}-{\cal K}_{21}{\cal K}_{12} < 0 \\
\end{equation}
which is that $\det[{\cal K}] < 0$, the same condition in equation (\ref{hi_ord_ins}). We subsequently use this slow enzyme assumption when dealing with enzymes and other interacting, fast components.

\subsection{On the required magnitude of the substrate-vs-product mobility asymmetry}

The instability condition, see Eq.~7 in the main text or (\ref{eq:inst}) above, can be rewritten as
\begin{equation}
	\frac{\Dpe-\Dse}{\Dpe} > \frac{ (\Rs+\Rp)^2\left(\frac{\vs}{\ve \phie}+\frac{1}{1-\phie}\right)}{\kspo\kcat(1-e^{-\beta \Delta \mu})+ (\Rs+\Rp)\Rp\left(\frac{\vs}{\ve \phie}+\frac{1}{1-\phie}\right)}. \label{eq:ratio}
\end{equation}
In general, we expect both the catalyzed and the spontaneous reaction to be strongly driven. Indeed, a typical value for $-\Delta \epsilon$ will be of the order of a few $k_\mathrm{B}T$, whereas for $\Delta \mu$ we would expect even larger values, a typical value for ATP hydrolysis being $\Delta \mu \approx 24 k_\mathrm{B}T$. In this strongly driven limit, we may set $e^{\beta \Delta \epsilon} \ll 1$, $e^{-\beta \Delta \mu} \ll 1$, and $e^{-\beta(\Delta \epsilon + \Delta \mu)} \ll 1$, and (\ref{eq:ratio}) simplifies to
\begin{equation}
	\frac{\Dpe-\Dse}{\Dpe} >  \frac{\frac{\kcat}{\kspo}\left(\frac{\kspo}{\kcat}+\phie\right)^2\left(\frac{\vs}{\ve}\frac{1}{\phie}+\frac{1}{1-\phie}\right)}{1+\left(\frac{\kspo}{\kcat}+\phie\right)\left(\frac{\vs}{\ve}\frac{1}{\phie}+\frac{1}{1-\phie}\right)}. \label{eq:ratio2}
\end{equation}
Eq.~\ref{eq:ratio2} gives the location of the spinodal region in the plane spanned by $\phie$ and the substrate/product mobility asymmetry $\frac{\Dpe-\Dse}{\Dpe}$ in the strongly-driven limit, which will vary depending on the choice of $\vs/\ve$ and $\kspo/\kcat$.

A typical enzyme is a large protein with a size of the order of $10$~nm, whereas a typical substrate is a small molecule with a size of the order of $0.5$~nm. We thus expect $\vs/\ve$ to be very small in realistic systems, of the order of $\vs/\ve \approx (10/0.5)^3 \approx 10^{-4}$. In this limit of $\vs/\ve \to 0$, the instability condition simplifies even further to
\begin{equation}
	\frac{\Dpe-\Dse}{\Dpe} >  \frac{\frac{\kcat}{\kspo}\left(\frac{\kspo}{\kcat}+\phie\right)^2}{1+\frac{\kspo}{\kcat}}. \label{eq:ratio3}
\end{equation}
Eq.~\ref{eq:ratio3} indeed provides a good approximation to (\ref{eq:ratio2}) for $\vs/\ve \approx 10^{-4}$, as can be checked numerically. In particular, it provides an accurate approximation of the location of the critical point, which is located at $\phie^\mathrm{crit} \to 0$ and
\begin{equation}
	\left(\frac{\Dpe - \Dse}{\Dpe}\right)^\mathrm{crit} = \frac{1}{1+\frac{\kcat}{\kspo}}.\label{eq:ratiocrit}
\end{equation}
Therefore, in the realistic limit of strongly driven reactions and a large size of the enzyme relative to the substrate size, CIPS is possible whenever
\begin{equation}
	\frac{\Dpe-\Dse}{\Dpe} > \frac{1}{1+\frac{\kcat}{\kspo}} \Longleftrightarrow \frac{\Dpe-\Dse}{\Dse} > \frac{\kspo}{\kcat} .
\end{equation}
Both inequalities are equivalent. Importantly, the right hand side of these inequalities gets arbitrarily small as $\kcat/\kspo$ increases. This implies that, if the kinetics of the catalyzed reaction are much faster than the kinetics of the spontaneous reaction, i.e.~$\kcat/\kspo \gg 1$, which is generally expected as this is indeed the definition of a catalyzed reaction, the substrate/product mobility asymmetry required for CIPS to occur can get arbitrarily small.

Interestingly, beyond this critical point, the spinodal region is determined by
\begin{equation}
	0<\phie<\sqrt{\frac{\kspo}{\kcat}\left(1+\frac{\kspo}{\kcat}\right)\left(\frac{\Dpe-\Dse}{\Dpe}\right)}-\frac{\kspo}{\kcat},
\end{equation}
as obtained from rearranging (\ref{eq:ratio3}). Thus, in the limit of strongly driven reactions and large enzyme size relative to substrate size, CIPS is expected to occur at low enzyme concentrations. Phase separation in this limit will lead to an enzyme rich phase ($\phie>0$) coexisting with a phase fully depleted of enzyme ($\phie=0$).

\section{Stability of the four-component system}

An additional solvent component (e.g.~water) can be added to the system similarly to the enzyme component, with conserved dynamics and no reaction terms. Following a similar procedure as for the no-solvent case, we can expand around a steady state where we now have $\phisp^* \neq 1 - \phi^*_\mathrm{E}$ but $\phiw^* = 1 - \phisp^* - \phi^*_\mathrm{E}$. We can eliminate the solvent volume fraction to give a $3\times3$ matrix describing the evolution of E, S, and P volume fractions
\begin{equation}
	(\dot{\delta\phie}, \dot{\delta\phis}, \dot{\delta\phip})^T = {\cal K}_{\rm w}(\delta\phie, \delta\phis, \delta\phip)^T    
\end{equation}
with
\newline
\noindent
\makebox[\textwidth]{\parbox{1.3\textwidth}{%
		\begin{equation}
			{\cal K}_{\rm w} = 
			\begin{tiny}
				\begin{pmatrix}
					(\Mse^*+\Mpe^*+\Mwe^*)\frac{\bm{q}^2}{\phie^*}+\Mwe^*\frac{\ve}{\vw}\frac{\bm{q}^2}{\phiw^*} & -\Mse^*\frac{\ve}{\vs}\frac{\bm{q}^2}{\phis^*}+\Mwe^*\frac{\ve}{\vw}\frac{\bm{q}^2}{\phiw^*} & -\Mpe^*\frac{\ve}{\vs}\frac{\bm{q}^2}{\phip^*}+\Mwe^*\frac{\ve}{\vw}\frac{\bm{q}^2}{\phiw^*}\\
					-\Mse^*\frac{\bm{q}^2}{\phie^*}-\Ree +\Mws^*\frac{\vs}{\vw}\frac{\bm{q}^2}{\phiw^*} & (\Mse^*\frac{\ve}{\vs}+\Msp^*+\Mws^*)\frac{\bm{q}^2}{\phis^*}-\Rs  +\Mws^*\frac{\vs}{\vw}\frac{\bm{q}^2}{\phiw^*}& -\Msp^*\frac{\bm{q}^2}{\phip^*}+\Rp +\Mws^*\frac{\vs}{\vw}\frac{\bm{q}^2}{\phiw^*}\\
					-\Mpe^*\frac{\bm{q}^2}{\phie^*}+\Ree +\Mwp^*\frac{\vs}{\vw}\frac{\bm{q}^2}{\phiw^*} & -\Msp^*\frac{\bm{q}^2}{\phis^*}+\Rs +\Mwp^*\frac{\vs}{\vw}\frac{\bm{q}^2}{\phiw^*} & (\Mpe^*\frac{\ve}{\vs}+\Msp^*+\Mwp^*)\frac{\bm{q}^2}{\phip^*}-\Rp +\Mwp^*\frac{\vs}{\vw}\frac{\bm{q}^2}{\phiw^*}\\
				\end{pmatrix}
			\end{tiny}
		\end{equation}
}}

We now assume that the substrate and product dynamics are fast in comparison to the enzyme. Setting $\dot{\delta\phis}=\dot{\delta\phip}=0$ gives
\begin{equation}
	\dot{\delta\phie} = ({\cal K}_{\rm w}^{-1})_{1, 1}\delta\phie
\end{equation}
and so the system is unstable when
\begin{equation}
	({\cal K}_{\rm w}^{-1})_{1, 1} = \frac{({\cal K}_{\rm w})_{2, 2}({\cal K}_{\rm w})_{3, 3}-({\cal K}_{\rm w})_{2, 3}({\cal K}_{\rm w})_{3, 2}}{\det({\cal K}_{\rm w})} \geq 0.
	\label{solvstab}
\end{equation}
The numerator of (\ref{solvstab}) is, to order $\bm{q}^2$,
\begin{equation}
	\bm{q}^2\Bigg[-\Rp \frac{\Mse^*\frac{\ve}{\vs}+\Mws^*}{\phis^*}-\Rs \frac{\Mpe^*\frac{\ve}{\vs}+\Mwp^*}{\phip^*}-\frac{\vs}{\vw\phiw^*}\big(\Rs +\Rp \big)\big(\Mws^*+\Mwp^*\big)\Bigg].
	\label{solvnum}
\end{equation}
With the common choice of mobilities, $M_{ij}<0$ for $i \neq j$, and $R_i \geq 0$, the expression (\ref{solvnum}) is positive and the instability condition is simply
\begin{equation}
	\det({\cal K}_{\rm w}) > 0,
\end{equation}
which marks the transition from having all negative eigenvalues to having two negative and one positive eigenvalue. This expression is used to derive the instability (spinodal) line in Fig.~2(d) of the main text.

The determinant of ${\cal K}_{\rm w}$ has many terms and even when truncated to order $\bm{q}^2$ it is not a manageable expression. A key observation is that in the low solvent limit ($\phiw \to 0$), and assuming Kramer's form for the mobilities, we recover the no-solvent case, as can be seen in Fig.~2(d) in the main text.

\section{Effective free energy and binodals for the three-component system}

The effective free energy density $f_\mathrm{eff}$  can be determined by integration of $\mu_\mathrm{eff}$, such that $f_\mathrm{eff}(\phie)=(1/\ve)\int\mu_\mathrm{eff}(\phie)\mathrm{d}\phie$. Since $\mu_\mathrm{eff}$ can be determined up to a constant term, $f_\mathrm{eff}$ can be determined up to addition of an affine function in $\phie$. The integration gives
\begin{equation}
	\begin{split}
		\frac{f_\mathrm{eff}(\phie)}{k_{\rm B} T} &= \frac{1}{\vp}\frac{\kspo}{\kcat}\Bigg(\frac{1+e^{\beta \Delta \varepsilon}}{1+e^{-\beta (\Delta \varepsilon + \Delta \mu)}}\log\left[\frac{\kcat}{\kspo}\phie + e^{\beta (\Delta \varepsilon + \Delta \mu)}(1+e^{\beta \Delta \varepsilon} + \frac{\kcat}{\kspo}\phie)\right] \\
		-& \frac{\Dse+\Dpe e^{\beta \Delta \varepsilon}}{\Dpe+\Dse e^{-\beta (\Delta \varepsilon + \Delta \mu)}}\log\Big[\Dse\frac{\kcat}{\kspo}\phie + e^{\beta (\Delta \varepsilon + \Delta \mu)}(\Dse+\Dpe e^{\beta \Delta \varepsilon} + \Dpe \frac{\kcat}{\kspo}\phie)\Big]\Bigg) \\
		+&\left(\frac{1}{\vp}-\frac{1}{\ve}\right)\phie + \frac{1}{\vp}\log[1-\phie] + \phie\bigg(\frac{1}{\ve}\log\phie - \frac{1}{\vp} \log[\Dse\phis^*(\phie)+\Dpe\phip^*(\phie)]\bigg).
	\end{split}
\end{equation}

The effective dynamics $\dot{\phie}\approx \bm{\nabla} \cdot (\Mee\bm{\nabla}\mu_\mathrm{eff})$, with $\mu_\mathrm{eff}=\ve f'_\mathrm{eff}(\phie)$, drive the system towards a state that minimizes the total effective free energy, under the constraints that the amount of enzyme and the volume of the system are conserved. In particular, the effective free energy of a system that has phase separated into two phases, with volumes $V^{(1)}$ and $V^{(2)}$, and enzyme concentrations $\phie^{(1)}$ and $\phie^{(2)}$, is given by $F_\mathrm{eff}=f_\mathrm{eff}(\phie^{(1)})V^{(1)}+f_\mathrm{eff}(\phie^{(2)})V^{(2)}$. This must be minimized while keeping a constraint on the amount of enzyme, $(\phie^{(1)}V^{(1)}+\phie^{(2)}V^{(2)})/\ve$, and the volume of the system, $V^{(1)}+V^{(2)}$. Using Lagrange multipliers for these two constraints which we denote (for reasons that will immediately become obvious) as $\mu_\mathrm{eff}$ and $p_\mathrm{eff}$, we must therefore minimize the function
\begin{equation}
	\mathcal{F}=f_\mathrm{eff}(\phie^{(1)})V^{(1)}+f_\mathrm{eff}(\phie^{(2)})V^{(2)} - \mu_\mathrm{eff} \frac{1}{\ve} \left( \phie^{(1)}V^{(1)}+\phie^{(2)}V^{(2)} \right) + p_\mathrm{eff} \left( V^{(1)}+V^{(2)} \right)
\end{equation}
relative to $\phie^{(i)}$ and $V^{(i)}$ for $i=1,2$. From $\partial \mathcal{F}/\partial\phie^{(i)}=0$, we obtain the condition
\begin{equation}
	\ve f'_\mathrm{eff}(\phie^{(1)}) = \mu_\mathrm{eff} = \ve f'_\mathrm{eff}(\phie^{(2)}) ,\label{eq:equalmu}
\end{equation}
whereas from $\partial \mathcal{F}/\partial V^{(i)}=0$, we obtain the condition
\begin{equation}
	f'_\mathrm{eff}(\phie^{(1)})\phie^{(1)} - f_\mathrm{eff}(\phie^{(1)}) = p_\mathrm{eff} = f'_\mathrm{eff}(\phie^{(2)})\phie^{(2)} - f_\mathrm{eff}(\phie^{(2)}). \label{eq:equalp}
\end{equation}

Eqs.~\ref{eq:equalmu} and \ref{eq:equalp} respectively correspond to the equal slope and equal intercept conditions in the graphical common tangent construction, and are used to obtain the analytical prediction for the binodal lines in Figs. 2(a) and (b) of the main text. Eqs.~\ref{eq:equalmu} and \ref{eq:equalp} could also be interpreted as saying that the two phases must have equal effective chemical potential $\mu_\mathrm{eff} = \ve f'_\mathrm{eff}(\phie)$, and equal effective thermodynamic pressure $p_\mathrm{eff}=f'_\mathrm{eff}(\phie)\phie - f_\mathrm{eff}(\phie)$. However, it is important to note that in reality we are dealing with a nonequilibrium system, and moreover the effective quantities above were only obtained in the limit of fast reactions. In particular, as in other nonequilibrium systems \cite{wittkowski2014scalar}, the thermodynamic pressure $p_\mathrm{eff}$ may not necessarily coincide with the mechanical pressure exerted by the system on the walls of its container.

\section{Connection to mass-conserving reaction-diffusion systems}

Reaction-diffusion systems are called mass-conserving when each individual component is not conserved, but the sum of all components (total mass) is. These systems have recently attracted attention as they allow for a description of their instabilities beyond the linear level \cite{brauns2020phase,brauns2021wavelength}. We describe here how the ternary system introduced in the main text may be described as a two-component reaction-drift-diffusion system. More precisely, the equations for the evolution of the substrate and product fields may be written as
\begin{align}
	\dot{\phis} &= \bm{\nabla} \cdot \big(\Mse\bm{\nabla}\mue + \Mss\bm{\nabla}\mus + \Msp\bm{\nabla}\mup) - R(\phis,\phip), \label{evoS2}\\
	\dot{\phip} &= \bm{\nabla} \cdot \big(\Mpe\bm{\nabla}\mue + \Mps\bm{\nabla}\mus + \Mpp\bm{\nabla}\mup) + R(\phis,\phip). \label{evoP2}
\end{align} 
Importantly, note that these two equations are sufficient to describe the full dynamics of the system, as the enzyme concentration is fully determined by the incompressibility constraint $\phie=1-\phis-\phip$.

The sum of substrate and product $\phi_\mathrm{s+p}=\phis+\phip$ is conserved, with dynamics given by
\begin{equation}
	\dot{\phi}_\mathrm{s+p} = \bm{\nabla} \cdot [ (\Mse+\Mpe)\bm{\nabla}\mue + (\Mss+\Mps)\bm{\nabla}\mus + (\Msp+\Mpp)\bm{\nabla}\mup ].
\end{equation}
Using the incompressibility condition and Onsager symmetry of the mobilities, this equation becomes
\begin{equation}
	\dot{\phi}_\mathrm{s+p} = \bm{\nabla} \cdot [ (\Mse+\Mpe)\bm{\nabla}\mue - (\ve/\vs) \Mse\bm{\nabla}\mus - (\ve/\vs)\Mpe\bm{\nabla}\mup ].
\end{equation}
Finally, using the form $M_{ij} = - \beta D_{ij} \phi_i \phi_j$ for the mobilities (with $i \neq j$), and $\bm{\nabla} \mu_j = \frac{k_\mathrm{B}T}{\phi_j}\bm{\nabla}\phi_j$ for the gradient of the chemical potentials, we can rewrite the evolution equation as
\begin{equation}
	\dot{\phi}_\mathrm{s+p} = - \bm{\nabla} \cdot [ \tilde{M}_\mathrm{ee}(\phis,\phip)  \bm{\nabla} \tilde{\mu}_\mathrm{eff}(\phis,\phip)],
	\label{eq:conserv}
\end{equation}
where we have defined
\begin{equation}
	\tilde{M}_\mathrm{ee}(\phis,\phip) \equiv \beta (\Dse\phis+\Dpe\phip) (1-\phis-\phip),
\end{equation}
\begin{equation}
	\tilde{\mu}_\mathrm{eff}(\phis,\phip) \equiv k_\mathrm{B} T [ \log(1-\phis-\phip) - (\ve/\vs)\log(\Dse \phis+\Dpe \phip)].
\end{equation}

Through the incompressibility condition $\phi_\mathrm{s+p}+\phie=1$, equation (\ref{eq:conserv}) implies an evolution equation for the enzyme volume fraction 
\begin{equation}
	\dot{\phi}_\mathrm{e} = \bm{\nabla} \cdot [ \tilde{M}_\mathrm{ee}(\phis,\phip)  \bm{\nabla} \tilde{\mu}_\mathrm{eff}(\phis,\phip)],
	\label{eq:conservE}
\end{equation}
which is identical in form to the equation for the effective enzyme evolution obtained in section \emph{Effective free energy and binodal} of the main text. However, here $\tilde{M}_\mathrm{ee}(\phis,\phip)$ and $\tilde{\mu}_\mathrm{eff}(\phis,\phip)$ are to be understood as functions of $\phis$ and $\phip$, whereas $\Mee(\phie)$ and $\mu_\mathrm{eff}(\phie)$ in the main text were to be interpreted as functions of $\phie$, as in the latter case $\phis$ and $\phip$ had been enslaved to $\phie$ through the approximation of fast local equilibration to the reaction equilibria $\phis^*(\phie)$ and $\phip^*(\phie)$.

Equation \ref{eq:conserv}, on the other hand, was obtained without any approximations, and allows us to draw significant parallels to the literature on mass-conserving reaction-diffusion systems \cite{brauns2020phase,brauns2021wavelength}. First, it shows that the total mass $\phi_\mathrm{s+p}$ evolves according to an effective potential, which in our case we identify with an effective chemical potential, while in Refs.~\citenum{brauns2020phase,brauns2021wavelength} it was termed the \emph{mass redistribution potential}. Second, it implies that, for a closed system at steady state, solutions must satisfy  $\tilde{\mu}_\mathrm{eff}(\phis,\phip)=\text{constant}$. In the framework of Refs.~\citenum{brauns2020phase,brauns2021wavelength}, this condition defines the \emph{flux balance subspace}. Finally, to obtain a closed system of equations, Eq.~\ref{eq:conserv} still needs to be complemented with either of the evolution equations for the substrate or product volume fractions (\ref{evoS2}) or (\ref{evoP2}). The condition of vanishing net reaction flux in these equations $R(\phis,\phip)=0$ constitutes the equivalent of the \emph{reaction nullcline} defined in Refs.~\citenum{brauns2020phase,brauns2021wavelength}.

In spite of these similarities, there are also a number of significant differences between the CIPS model and mass-conserving reaction diffusion systems: (i) the mass redistribution potential and the flux balance subspaces in Refs.~\citenum{brauns2020phase,brauns2021wavelength} are linear in the concentrations, whereas the equivalents here are non-linear; (ii) the coefficient governing the response to gradients of the mass redistribution potential is constant in Refs.~\citenum{brauns2020phase,brauns2021wavelength}, whereas here $\tilde{M}_\mathrm{ee}(\phis,\phip)$ is concentration-dependent; (iii) the conserved part of the dynamics of the non-conserved quantities in Refs.~\citenum{brauns2020phase,brauns2021wavelength} is purely self-diffusive, whereas here we have cross-diffusive (drift) components as well.

Interestingly, two-component mass-conserving reaction-diffusion systems have been shown to always undergo uninterrupted coarsening following an instability, that is, these systems always show macrophase separation rather than finite-wavelength pattern formation (or microphase separation) \cite{brauns2021wavelength}. This is in line with what we find here for CIPS, for which we have only observed uninterrupted coarsening leading to macrophase separation in numerical solutions of the full evolution equations (in addition to other indirect evidence for macrophase separation such as the linear instability analysis showing a long wavelength $q^2 \to 0$ instability, or the existence of an effective free energy in the limit of fast equilibration of the reaction fluxes). Future work may further explore the connection of CIPS-like models that use a thermodynamically consistent dynamics following from a free energy, and mass-conserving systems with ad-hoc reaction and diffusion terms.

\section{Connection to cross-diffusion and phoresis}

\subsection{General derivation}

The Onsager framework of nonequilibrium thermodynamics used in the main text can also be reinterpreted in terms of Fickian cross-diffusion dynamics \cite{vanag2009cross}, or similarly, in terms of diffusive dynamics with an additional phoretic drift \cite{anderson1989colloid,golestanian2019phoretic}. We start from the general theory of conserved dynamics of an incompressible mixture described in the main text. The evolution equation for each component, with $i=1,...,N$, reads
\begin{equation}
	\dot{\phi}_i = \bm{\nabla}\cdot \left[ \sum_{j=1}^N M_{ij}\bm{\nabla}\mu_j \right]
	\label{eq:evolM}
\end{equation}
where the chemical potential is given by $\mu_j=v_j\frac{\delta F}{\delta \phi_j}$, and the mobilities satisfy the Onsager symmetries $v_j M_{ij}=v_i M_{ji}$ as well as the incompressibility condition $M_{jj}= - \sum_{i\neq j} M_{ij}$. Note that using these two conditions on the mobilities we can rewrite (\ref{eq:evolM}) as
\begin{equation}
	\dot{\phi}_i = \bm{\nabla}\cdot \left[ \sum_{\substack{j=1 \\ j\neq i}}^N M_{ij}\bm{\nabla} \left(\mu_j - \frac{v_j}{v_i}\mu_i\right) \right].
	\label{eq:evolM2}
\end{equation}

In order to turn (\ref{eq:evolM}) into a Fickian (cross-)diffusion equation, we must (i) single out a given component as the solvent, say component $N$ without loss of generality, which is then eliminated through $\phi_N=1 - \sum_{i=1}^{N-1} \phi_i$; (ii) rewrite the dynamics in terms of concentrations $c_i$ rather than volume fractions $\phi_i$, where $c_i=\phi_i/v_i$; and (iii) rewrite the fluxes as being driven by concentration gradients rather than chemical potential gradients, by applying the chain rule. In this way, (\ref{eq:evolM2}) can be shown to be equivalent to
\begin{equation}
	\dot{c}_i = \bm{\nabla}\cdot \left[ \sum_{k=1}^{N-1} \Delta_{ik}\bm{\nabla} c_k \right]
	\label{eq:evolc}
\end{equation}
which now applies only to the non-solvent components $i=1,...,N-1$, and where we have defined the diffusion tensor
\begin{equation}
	\Delta_{ik} \equiv \frac{v_k}{v_i} \sum_{\substack{j=1 \\ j\neq i}}^N M_{ij} \frac{\partial}{\partial \phi_k}\left(\mu_j - \frac{v_j}{v_i}\mu_i\right).
	\label{eq:Delta}
\end{equation}
Crucially, we note that the chemical potentials $\mu_{i,j}$ in (\ref{eq:Delta}) should be interpreted as functions of the non-solvent components $\mu_{i,j}(\{\phi_1,...,\phi_{N-1}\})$.

Equation~(\ref{eq:Delta}) shows that the self- and cross-diffusion coefficients generally involve both a purely kinetic or frictional part (governed by the Onsager mobilities $M_{ij}$), as well as an energetic contribution given by the derivatives of the chemical potentials. The self-diffusion coefficients ($\Delta_{ii}$) should be related to the standard (single-particle) diffusion coefficient in the limit of vanishing $\phi_i$'s for all components other than the solvent. The cross-diffusion coefficients ($\Delta_{ij}$ with $i \neq j$) in turn describe the response of component $i$ to gradients in the concentration of component $j$, and as such are related to the phenomenon of phoresis. In particular, we may interpret from the form of (\ref{eq:evolc}) that component $i$ develops a drift velocity $\bm{V}_{ij} = - (\Delta_{ij}/c_i) \bm{\nabla}c_j$ in response to gradients of component $j$. If we define a phoretic mobility $\mu_{ij}$ as the coefficient entering the relation $\bm{V}_{ij} = - \mu_{ij} \bm{\nabla}c_j$, we find that the phoretic mobility is related to the cross-diffusion coefficient as
\begin{equation}
	\mu_{ij} = \frac{v_i \Delta_{ij}}{\phi_i}.
\end{equation}
This phoretic mobility, like the corresponding cross-diffusion coefficient, is positive or negative if component $i$ tends to move towards regions of lower or higher concentration of component $j$, respectively.

\subsection{General ternary mixture}

As a simple example, consider a mixture of two solutes (components $1$ and $2$) in a solvent ($3$). The self-diffusion coefficient of component 1 will be
\begin{equation}
	\Delta_{11} =   M_{12} \frac{\partial}{\partial \phi_1}\left(\mu_2 - \frac{v_2}{v_1}\mu_1\right)
	+ M_{13} \frac{\partial}{\partial \phi_1}\left(\mu_3 - \frac{v_3}{v_1}\mu_1\right)
	\label{eq:Delta11}
\end{equation}
and the cross-diffusion coefficient describing the response of 1 to gradients in the concentration of 2 will be
\begin{equation}
	\Delta_{12} =  \frac{v_2}{v_1} \left[ M_{12} \frac{\partial}{\partial \phi_2}\left(\mu_2 - \frac{v_2}{v_1}\mu_1\right)
	+ 		M_{13} \frac{\partial}{\partial \phi_2}\left(\mu_3 - \frac{v_3}{v_1}\mu_1\right) \right].
	\label{eq:Delta12}
\end{equation}
The corresponding coefficients $\Delta_{22}$ and $\Delta_{21}$ governing the dynamics of component 2 are obtained simply by exchanging $1 \leftrightarrow 2$.

\subsection{Ideal ternary mixture}

For an ideal ternary mixture, with only entropic contributions to its free energy, we have $\mu_1 = k_\mathrm{B}T \log \phi_1$, $\mu_2 = k_\mathrm{B}T \log \phi_2$, and $\mu_3 = k_\mathrm{B}T \log (1 - \phi_1 - \phi_2)$. The (cross-)diffusion coefficients above become
\begin{equation}
	\Delta_{11} = - \frac{k_\mathrm{B}T}{v_1} \left( \frac{M_{12}v_2 + M_{13}v_3}{\phi_1} + \frac{M_{13}v_1}{1-\phi_1-\phi_2}   \right),
	\label{eq:Delta11n}
\end{equation}
\begin{equation}
	\Delta_{12} =  - k_\mathrm{B}T \frac{v_2}{v_1} \left( - \frac{M_{12}}{\phi_2} + \frac{M_{13}}{1-\phi_1-\phi_2}   \right).
	\label{eq:Delta12n}
\end{equation}

\subsection{Ideal ternary mixture with Kramer-type mobilities}

We now use the Kramer form of the mobilities, with $M_{ij} = - \beta D_{ij} \phi_i \phi_j$ for $i\neq j$. The self-diffusion coefficient becomes
\begin{equation}
	\Delta_{11} = \frac{1}{v_1} \left[ D_{13}v_3(1-\phi_1-\phi_2) + D_{13}v_1\phi_1 + D_{12} v_2 \phi_2  \right]
	\label{eq:Delta11nk}
\end{equation}
or alternatively, using the Onsager symmetries, 
\begin{equation}
	\Delta_{11} =  D_{31}(1-\phi_1-\phi_2) + D_{13}\phi_1 + D_{21} \phi_2 .
	\label{eq:Delta11nkb}
\end{equation}

The cross-diffusion coefficient, on the other hand, becomes
\begin{equation}
	\Delta_{12} = \frac{v_2}{v_1}(D_{13} - D_{12}) \phi_1
	\label{eq:Delta12nkb}
\end{equation}
and therefore the corresponding phoretic mobility is simply
\begin{equation}
	\mu_{12} = v_2 (D_{13} - D_{12})
	\label{eq:mu12}
\end{equation}
which is a constant, independent of the concentrations of components 1 and 2.

\subsection{Dilute ideal ternary mixture with Kramer-type mobilities}

In the dilute limit with $\phi_1 \to 0$ and $\phi_2 \to 0$, we find for the self-diffusion coefficient
\begin{equation}
	\Delta_{11} \approx \frac{v_3}{v_1} D_{13} =  D_{31} \equiv D_{1}^{(0)}
	\label{eq:Delta11nkc}
\end{equation}
where $D_1^{(0)}$ can be interpreted as the standard diffusion coefficient of component 1 in the solvent.

Using (\ref{eq:mu12}) and (\ref{eq:Delta11nkc}), we see that we can derive a value for the off-diagonal Onsager coefficient $D_{12}$ if we have experimental measurements of the single-particle diffusion coefficient $D_{1}^{(0)}$ of component 1 in a solvent, and the phoretic mobility $\mu_{12}$ representing the response of component 1 to gradients in component 2 in the same solvent, through the simple relation
\begin{equation}
	D_{12} = \frac{v_1}{v_3} D_1^{(0)} - \frac{1}{v_2}\mu_{12}.
	\label{eq:expD12}
\end{equation}

\subsection{Consequences for CIPS}

The form of (\ref{eq:mu12}) is consistent with the requirement $\Dpe>\Dse$ for an instability to occur in the ternary system derived in the main text. If we take component 1 to represent the enzyme, component 2 to represent the substrate, and component 3 (solvent) to represent the product, (\ref{eq:mu12}) becomes $\mu_\mathrm{es} = \vs (\Dep - \Des) = \ve (\Dpe-\Dse)$, and the condition $\Dpe>\Dse$ implies $\mu_\mathrm{es}>0$, i.e.~the enzymes move towards lower substrate concentration. Because the enzymes consume substrate, this results in an effective enzyme-enzyme attraction.
Alternatively, we may take component 2 to represent the product and component 3 (solvent) to represent the substrate. Now, (\ref{eq:mu12}) becomes $\mu_\mathrm{ep} = \vs (\Des - \Dep) = \ve (\Dse-\Dpe)$, and the condition $\Dpe>\Dse$ implies $\mu_\mathrm{ep}<0$, i.e.~the enzymes move towards higher product concentration. Because the enzymes generate product, this again results in an effective enzyme-enzyme attraction.

Using (\ref{eq:expD12}) together with the Onsager symmetries allows us to write $\Dse = D_\mathrm{e}^{(0)}-\mu_\mathrm{es}/\ve$ and $\Dpe = D_\mathrm{e}^{(0)}-\mu_\mathrm{ep}/\ve$. These expressions could be used (assuming an ideal mixture) to extract the values of the relevant off-diagonal Onsager coefficients from measurements of the enzyme diffusion coefficient in water, as well as enzyme phoretic mobilities in response to substrate and product gradients, also in water solvent. Additionally, the difference $\Dpe-\Dse$, which plays a key role in the instability condition (Eq.~7 in the main text) can in this case be written as
\begin{equation}
	\Dpe-\Dse = \frac{\mu_\mathrm{es} - \mu_\mathrm{ep}}{\ve}
	\label{eq:diff}
\end{equation}
which again is consistent with the necessary condition for instability $\Dpe-\Dse>0$, as it implies $\mu_\mathrm{es} - \mu_\mathrm{ep}>0$, i.e.~the enzyme must be more repelled from (or less attracted to) the substrate than the product.

Importantly, measurements for a number of enzymes \cite{agudo2018enhanced,zhang2021chemically} have shown significantly different drifts of the enzyme in response to gradients of the substrate and the product, implying $\mu_\mathrm{es} \neq \mu_\mathrm{ep}$, which would in turn imply $\Dpe \neq \Dse$ within the model described here. However, the importance of energetic effects relative to kinetic effects in these experiments (i.e.~the ideality of the enzyme-substrate-product mixture) in these experiments is as of yet unclear. Similarly, the underlying mechanism of synthetic phoretic microswimmers relies on the existence of a significant difference in the phoretic response of the swimmer to the substrate and product of a catalyzed reaction \cite{golestanian2019phoretic}.

\section{Connection to Maxwell-Stefan diffusion}

An alternative framework for multicomponent diffusion, which is equivalent to both the Onsager non-equilibrium thermodynamic framework used in the main text and the cross-diffusion framework presented above, is the Maxwell-Stefan model \cite{taylor1993multicomponent}. In this case, one establishes again a connection between the chemical potential gradients, which act as driving forces, and the fluxes of the different species. However, in contrast with the Onsager formalism, one postulates that the thermodynamic force on component $i$ is balanced by a frictional force arising from the difference in the velocity of $i$ and the velocities of all other components. Here, the velocity $\bm{V}_j$ of a component $j$ is related to the concentration flux $\bm{J}^c_j$ of component $j$ by $\bm{V}_j=\bm{J}^c_j/c_j$, where $c_j$ is the concentration. This force balance can be written as
\begin{equation}
	\frac{\bm \nabla \mu_i}{k_\mathrm{B}T} = \sum_{\substack{j=1 \\ j\neq i}}^N \frac{c_j}{c_\mathrm{tot} \mathcal{D}_{ij}}\left(\frac{\bm{J}^c_j}{c_j}-\frac{\bm{J}^c_i}{c_i}\right)
	\label{eq:MS} 
\end{equation}
where $c_\mathrm{tot}=\sum_{i=1}^N c_i$. The coefficients $\mathcal{D}_{ij}$, which are symmetric under exchange of $i$ and $j$ and are only defined for $i \neq j$, are known as the Maxwell-Stefan diffusivities. The Maxwell-Stefan diffusivities represent an inverse friction coefficient between components $i$ and $j$. With the constraints just described, we notice that there are $N(N-1)/2$ independent Maxwell-Stefan diffusivities, just like there are $N(N-1)/2$ independent Onsager mobilities $M_{ij}$.

Indeed, it is possible to invert equation (\ref{eq:MS}) and thus obtain a direct link between the Maxwell-Stefan diffusivities and the Onsager mobilities, although this becomes a cumbersome procedure for $N>2$ \cite{taylor1993multicomponent}. In any case, one finds that the Onsager mobilities are a combination of the Maxwell-Stefan diffusivities (themselves concentration-dependent in general) and the concentrations of the different components. Unlike the Fickian cross-diffusion coefficients, the Maxwell-Stefan diffusivities are, like the Onsager mobilities, of purely kinetic or frictional origin, i.e.~they do not depend on the form of the free energy.

The Maxwell-Stefan diffusivities have been measured both experimentally and in molecular dynamics simulations for a multitude of mixtures \cite{taylor1993multicomponent,guevara2016mutual,guevara2018interplay}. It is found that even minor changes in the molecular structure (e.g.~shape, polarity, etc.) of the components of the mixture can lead to substantial changes in the Maxwell-Stefan diffusivities, which translate into substantial changes in the Onsager coefficients. As an example, ternary mixtures of acetone-benzene-methanol show widely different Maxwell-Stefan diffusivities and Onsager coefficients than mixtures of acetone-benzene-ethanol, even if the molecular differences between ethanol and methanol may appear to be minor \cite{guevara2016mutual}. Maxwell-Stefan diffusion has also been shown to be capable of driving motion of DNA origami particles in response to oscillatory pattern formation of the MinDE protein system \emph{in vitro} \cite{ramm2021diffusiophoretic}.

\end{document}